\documentclass[manuscript,screen]{acmart}
\usepackage{subcaption}
\usepackage{graphicx} 
\AtBeginDocument{%
  \providecommand\BibTeX{{%
    \normalfont B\kern-0.5em{\scshape i\kern-0.25em b}\kern-0.8em\TeX}}}
    
\setcopyright{acmcopyright}
\copyrightyear{2022}
\acmYear{2022}
\acmDOI{XXXXXXX.XXXXXXX}

\acmConference[ICMI '22]{}{Nov. 7--11, 2022}{Bengaluru, IN}
    
\begin{document}
\def\x{{\mathbf x}}
\def\L{{\cal L}}
\def\eg{\textit{e.g.}}
\def\ie{\textit{i.e.}}
\def\Eg{\textit{E.g.}}
\def\etal{\textit{et al.}}
\def\etc{\textit{etc}}

\title{Neural Encoding of Songs is Modulated by Their Enjoyment}


\author{Gulshan Sharma}
\email{gulshan.19csz0004@iitrpr.ac.in}
\orcid{0000-0002-5332-7256}
\affiliation{%
 \institution{Indian Institute of Technology Ropar}
  \country{India}
}

\author{Pankaj Pandey}
\orcid{0000-0003-4821-7068}
\affiliation{%
 \institution{Indian Institute of Technology Gandhinagar}
 \country{India}
}

\author{Ramanathan Subramanian}
\orcid{0000-0001-9441-7074}
\affiliation{%
 \institution{University of Canberra}
 \country{Australia}
}

\author{Krishna. P. Miyapuram}
\orcid{0000-0001-5779-2342}
\affiliation{%
 \institution{Indian Institute of Technology Gandhinagar}
 \country{India}
}

\author{Abhinav Dhall}
\orcid{0000-0002-2230-1440}
\affiliation{%
 \institution{Indian Institute of Technology Ropar}
 \country{India}
}

\begin{abstract}

We examine user and song identification from neural (EEG) signals. Owing to perceptual subjectivity in human-media interaction, music identification from brain signals is a challenging task. We demonstrate that subjective differences in music perception aid user identification, but hinder song identification. In an attempt to address intrinsic complexities in music identification, we provide empirical evidence on the role of \textit{enjoyment} in song recognition. Our findings reveal that considering song enjoyment as an additional factor can improve EEG-based song recognition.
\end{abstract}



\begin{teaserfigure}
  \includegraphics[width=\linewidth]{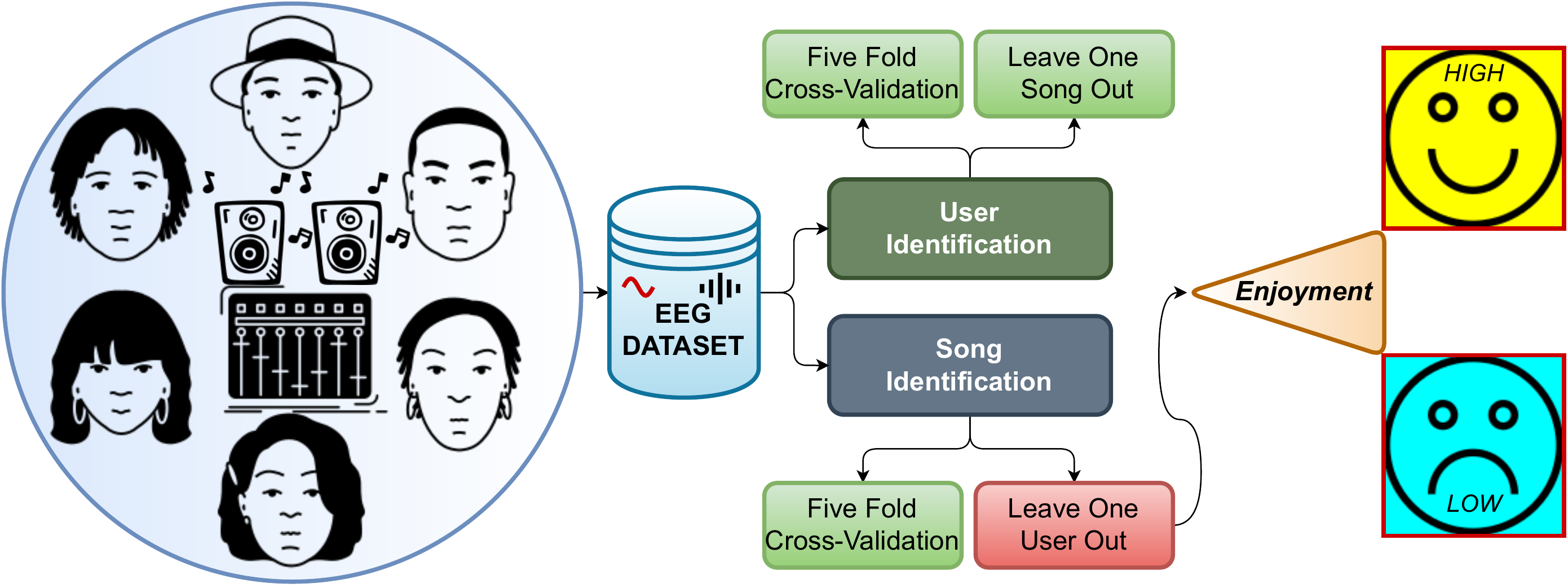}
  \caption{\textbf{Study Overview:} Users listen to songs and their EEG responses are recorded. We then employ EEG signals to perform (a) user and (b) song identification. Recognition results obtained with five-fold vs leave-one-out cross validation are discussed. Specifically, leave-one-user out song recognition substantially benefits from information concerning song enjoyment.}
  \label{fig:icmi_main}
\end{teaserfigure}

\maketitle

\section{Introduction \& Background}
Music is one of humanity's universal modes of expression and communication. People of all ages and cultures exhibit music appreciation, which is generally shaped by one's ambience and experiences. Music listening is a hedonistic activity, but its impact goes beyond mere entertainment; it influences our mood and creates a sense of calm in our lives. It has long been a strong motivator in fostering harmonious relationships between cultures and societies~\cite{large2010neurodynamics}. Naturalistic music is comprised of numerous compositional elements such as rhythm, melody, timbre, dynamics, \etc. These compositional elements generate a variety of computational features that are distributed across time and frequency domains~\cite{nawaz2018effect}. The absorption of music triggers several oscillatory systems in the brain. The brain can resonate with music's oscillatory properties, and this synchronization phenomenon is termed as neural entrainment \cite{doelling2015cortical,cirelli2016measuring}.

Neural entrainment enables us to differentiate between the brain's responses to two different musical perceptions. Even if two people listen to the same song, their brain's oscillatory systems may be activated differently, accounting for differences in perceptual and affective elements. Several studies have demonstrated various aspects of neural entrainment, such as beat and pitch recognition\cite{pandey2021predicting,tierney2014neural,nozaradan2014exploring}. A recent study discusses differences in music perception between musicians and non-musicians, and reports that audiovisual perception may be broadly correlated with auditory perception. Differences between musicians and non-musicians indicate musical experience as a specific factor influencing audiovisual perception~\cite{sorati2020audiovisual}. Thus, perception's role may entail bringing diverse signatures to the oscillatory activity of the brain. Musical aspects may be perceived differently depending on the listener's psychological state.\\
The effect of music on brain activations has been widely studied in multifarious contexts. EEG and fMRI are the two prevalent techniques utilized in neuroscientific brain analysis; EEG research is often conducted in an empirical setting involving a well-defined cognitive task. With advances in neurotechnology, the availability of a high-density MEG~\cite{DECAF15} and EEG~\cite{DEAP12,Ascertain17} systems enables analysis of complex brain interactions. Brain waves induced during music listening stimulate emotional and several higher-order cognitive processes \cite{de2020music,daly2020neural}. Based on morphological and functional aspects, these waves are organized into multiple frequency bands. The $\delta$ (1-3 Hz), $\theta$ (3-8 Hz), and $\alpha$ (8-13 Hz) bands are associated with sleep or relaxation, whereas the $\beta$ (13 - 30 Hz) and $\gamma$ (>30 Hz) bands are associated with attention and perception~\cite{bhattacharya2001long,Bilalpur18}. Hence, decomposing brain waves into frequency bands provides numerous perceptive insights, which can be further uncovered via state-of-the-art machine learning (ML) techniques~\cite{DEAP12, DECAF15}.
Multivariate EEG channels provide extensive data, and manually selecting the relevant feature for a specific task can be time-consuming. With the advancement in deep learning architectures, learning salient features enables solving complex classification problems.

Song identification from brain responses is a complex problem due to differences in music perception. Song identification based on brain activity is proposed in~\cite{sonawane2021guessthemusic}; the authors use two approaches, one in the time domain and another in the frequency domain. They consider the 2D image of the 1s EEG time response (\ie, channels $\times$ time points), and FFT decomposition of a 1s spectral frame (\ie, channels $\times$ Nyquist frequency) in the frequency domain. These images are fed into a convolutional neural network (CNN) for 5-fold and leave-one-user-out song classification. Their findings indicate poor time-domain results, with about chance-level accuracy for 5-fold cross validation-based song recognition. Frequency domain performance is superior, achieving $\approx 85\%$ accuracy. These findings reveal that identifiable spectral patterns are generated during music entrainment, and show the promise of spectral features for song recognition. Leave-one-user-out song recognition however produced chance-level accuracies in the time and spectral domains. 

Overall, the model performs poorly when the test user's data are \textit{left out} from the train set, highlighting subjectivity differences in music perception. One possible explanation is that people focus on distinct tones and singers during music entrainment, increasing variability across participants and decreasing performance of cross-participant song recognition. A recent article~\cite{pandey2022music} reports song recognition from EEG snippets corresponding to initial song segments on two public datasets. Here again, 5-fold classification results were much superior to leave-one-out classification, which was close to chance-level. 

Correlations between neural signatures and attributes such as enjoyment and familiarity during music listening are discussed in~\cite{lawhatre2020classifying} and~\cite{2017familiarity}. Previous research has attempted to classify liked versus disliked songs using various combinations of features from the time-frequency of EEG signal and demonstrated an improvement in accuracy when including the listener's familiarity in a feature vector \cite{hadjidimitriou2012toward,hadjidimitriou2013eeg}. However, no attempt has been made to classify the songs based on enjoyment rating.

In this regard, our study is the first to demonstrate an improvement in accuracy by accounting for enjoyment ratings. We make the following research contributions: (a)\textbf{ User Identification:} We empirically observe significant differences in EEG encodings across users, primarily owing to large perceptual differences during music listening. (b) \textbf{Song Identification:} We demonstrate the hardship in identifying a song's EEG encodings for a novel user. (c) \textbf{Effects of musical Appraisal on Song Identification:} Our novelty lies in employing \textit{enjoyment} ratings for song identification. While cross-subject (leave-one-user-out) EEG-based song classification leads to chance level recognition, segmenting data conditioned on enjoyment ratings enable better song recognition. This finding conveys that additional attributes such as enjoyment need to be considered for effective cross-user song recognition. Related experiments are detailed below.

\section{Experiments}
According to the identical and independent (i.i.d) assumption statistical learning, a model will not generalize well if the train and test data distributions differ. We investigate song and user identification using neural (EEG) recordings. We hypothesize that different songs and users have distinct neural encodings that reflect song and participant-related differences. As a result, we model our problem as a classification task, and we explore if a CNN-based model can learn appropriate encodings from these EEG recordings. We're also interested to see if these neural encodings are robust to unseen data. We validate this classification hypothesis using two approaches: (a) five-fold cross-validation, which divides the data into complementary subsets and tests whether the model is well generalized across the entire dataset. (b) To ensure an unbiased estimate of model performance, we also employ a leave-one-out validation scheme, in which a user's data as a whole is removed during model training, and model testing is performed on this hold-out data. This validation seeks to determine whether the model can generalize its learning to completely unseen data. Following subsections describe the user and song identification experiments.

\subsection{Dataset}
We examined a publicly available dataset NMED-T \cite{steven_losorelli_2017_1417917}, collected for music processing research. The dataset contains electroencephalogram (EEG) recordings and behavioral responses from twenty participants listening to ten full-length naturalistic songs. During dataset acquisition, songs were presented in random order. Following each trial, participants rated their familiarity and enjoyment of the song on a scale of 1–9. The EEG experiment was divided into two serial recording blocks to decrease participant fatigue and allow electrode impedance testing between recordings. We primarily used the pre-processed version of this dataset, which contains 125 channels of EEG data recorded at 125 Hz. Interested readers may refer to \cite{steven_losorelli_2017_1417917} for NMED-T description.

\subsection{Participant Identification} 
Our goal with EEG based-user identification is to investigate if individual user differences are discriminative when naturalistic songs are used as a stimulus. We look to learn a latent space embedding that preserves individual differences between participants. For 5-fold cross validation, we include all of the data for model training. This experiment reveals individual perceptual differences but does not reveal whether they are dependent or independent of the song class. Hence, we used the leave-one-song-out cross-validation strategy, in which we removed data samples of all users corresponding to one song iteratively, and trained the model on the remaining songs. We then tested the model on the excluded song data. The rationale behind this approach is to test the generalizability of latent space features for participant classification.

\subsection{Song Identification} 
Our brain processes millions of different sounds every day. These auditory stimuli travel from the ear to various cortical regions via auditory nerves, causing a variety of complex behavioral and emotional changes. These changes result in specific spatial activations across the cortex; hence a relation between the auditory stimuli and neural responses can be captured using a statistical model. Here, we aim to investigate the existence of discriminating latent neural embeddings across the song classes. Again we employ two approaches for song recognition. In the five-fold cross validation approach, We include all of the user data in the training phase. In the leave-one-user-out approach, we discard all song samples associated with a given participant and train the model on the remaining participants. The goal is to see if the model can capture generalized EEG song encodings across users. 

\begin{figure}[h]
      {\includegraphics[scale=0.4]{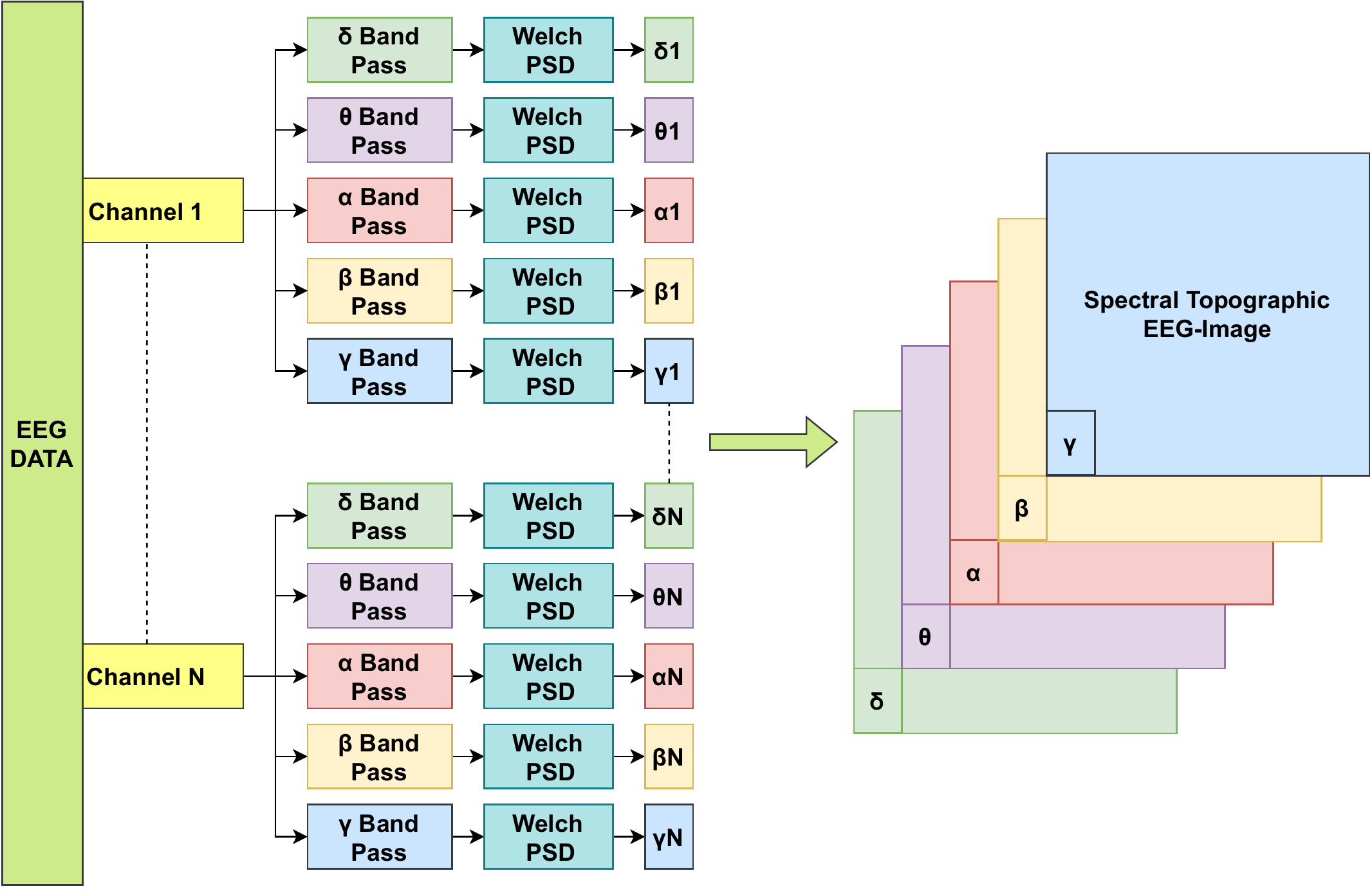}}
\vspace*{\fill}
\hspace*{\fill}
      {\includegraphics[scale=0.55]{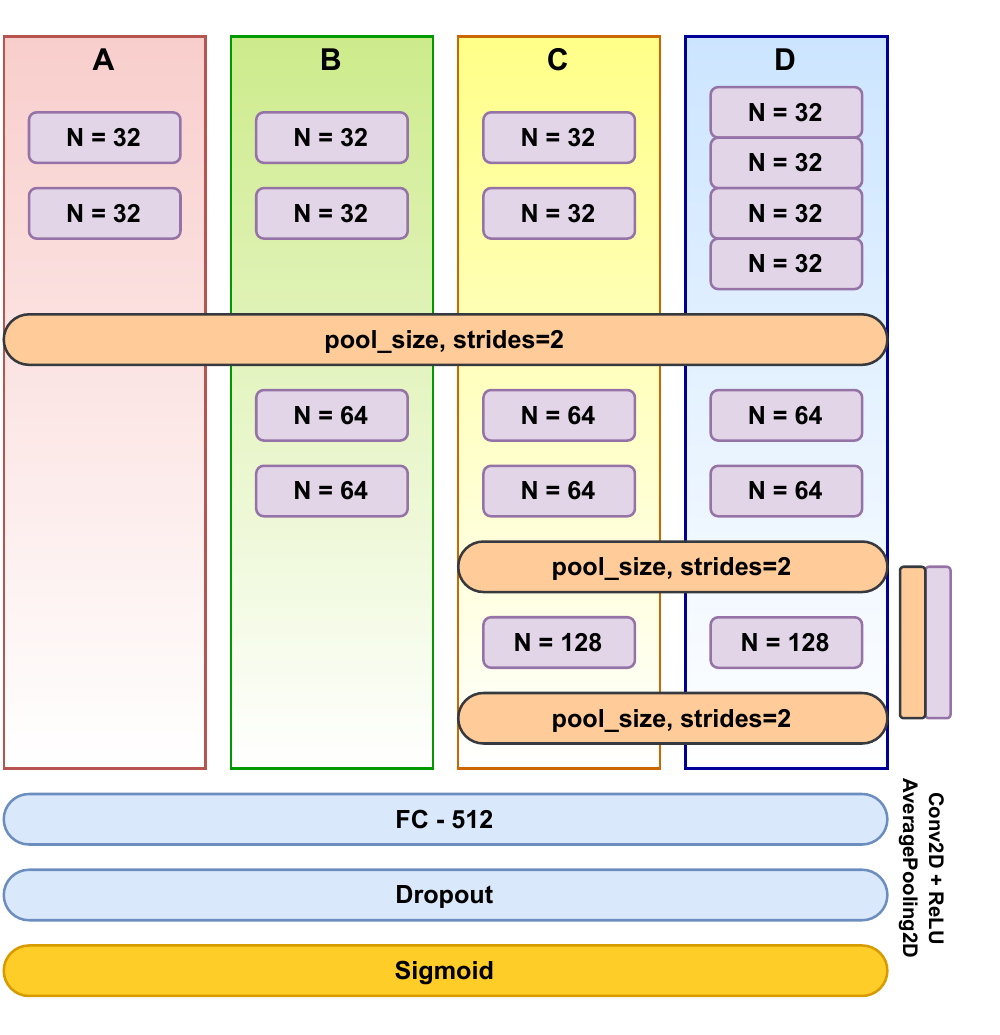}}
\caption{(a) Spectral Topographic Image Generation (Left), (b) CNN Architectures (Right)}
\label{feat-2dcnn}
\end{figure}

\subsection{Spectral Topographic Maps \& CNN Architectures}
EEG data contains multiple channels that register electrical activity from various cortical regions. For our investigation, we created spectral topographic maps \cite{Pouya} to encode the spatial and spectral information of the preprocessed EEG data. A 6$^{th}$ order Butterworth bandpass filter is applied to extract the $\delta$ (1-3 Hz), $\theta$ (3-8 Hz), $\alpha$ (8-13 Hz), $\beta$ (13-30 Hz), and $\gamma$ (>30 Hz) frequency bands. Welch's method is applied after bandpassing to estimate the power spectral density for each EEG channel. These estimates of power spectral density are then converted into a two-dimensional topographic image. Azimuthal equidistant projections are applied to preserve the inter-space between neighboring electrodes, and in-between electrode measurements are estimated using the Clough-Tocher interpolation scheme. These two-dimensional topographic images from each band are then concatenated to create a 5-dimensional spectral topographic map. Figure \ref{feat-2dcnn} (left) depicts the overall feature extraction strategy. To analyze these spectral topographic maps, we adopted four CNN architectures demonstrated in \cite{Pouya}. Figure \ref{feat-2dcnn} (right) depicts configuration of these architectures (A--D). Empirically, we found model A to work best for our data and all following results are presented for Model A. 

\section{Experimental Results \& Observations}
We begin our investigation with spectral topographic images extracted from 5-second EEG chunks. The initial goal was to empirically uncover the best-performing CNN architecture to identify participants and songs. All CNN architectures performed similarly well, but we chose architecture A due to its low complexity. For participant and song classification, we used ten repetitions of stratified 5-fold cross-validation and leave one-out cross-validation. Stratified guarantees that each per-class sample proportions in each fold is consistent with the entire dataset. In addition to the model test accuracy, we calculated weighted averaged test precision, recall, and F1-score.

\subsection{Participant Identification}

After ten repetitions of stratified 5-fold validation we obtained an average test accuracy, precision, recall and F1-score of $99.89\pm0.02, 99.89\pm0.01, 99.89\pm0.01$, and $99.89\pm0.01$ respectively. Observing ten repetitions of leave-one-song-out cross-validation, We found that the CNN model could effectively identify users on unseen songs. The model showed an average test accuracy, precision, recall and F1-score of $99.60 \pm 0.00, 99.62 \pm 0.00, 99.58 \pm 0.00$, and $99.58 \pm 0.00$ respectively.


\subsection{Song Identification}
\begin{figure}[h]
  \centering
  \includegraphics[scale=0.35]{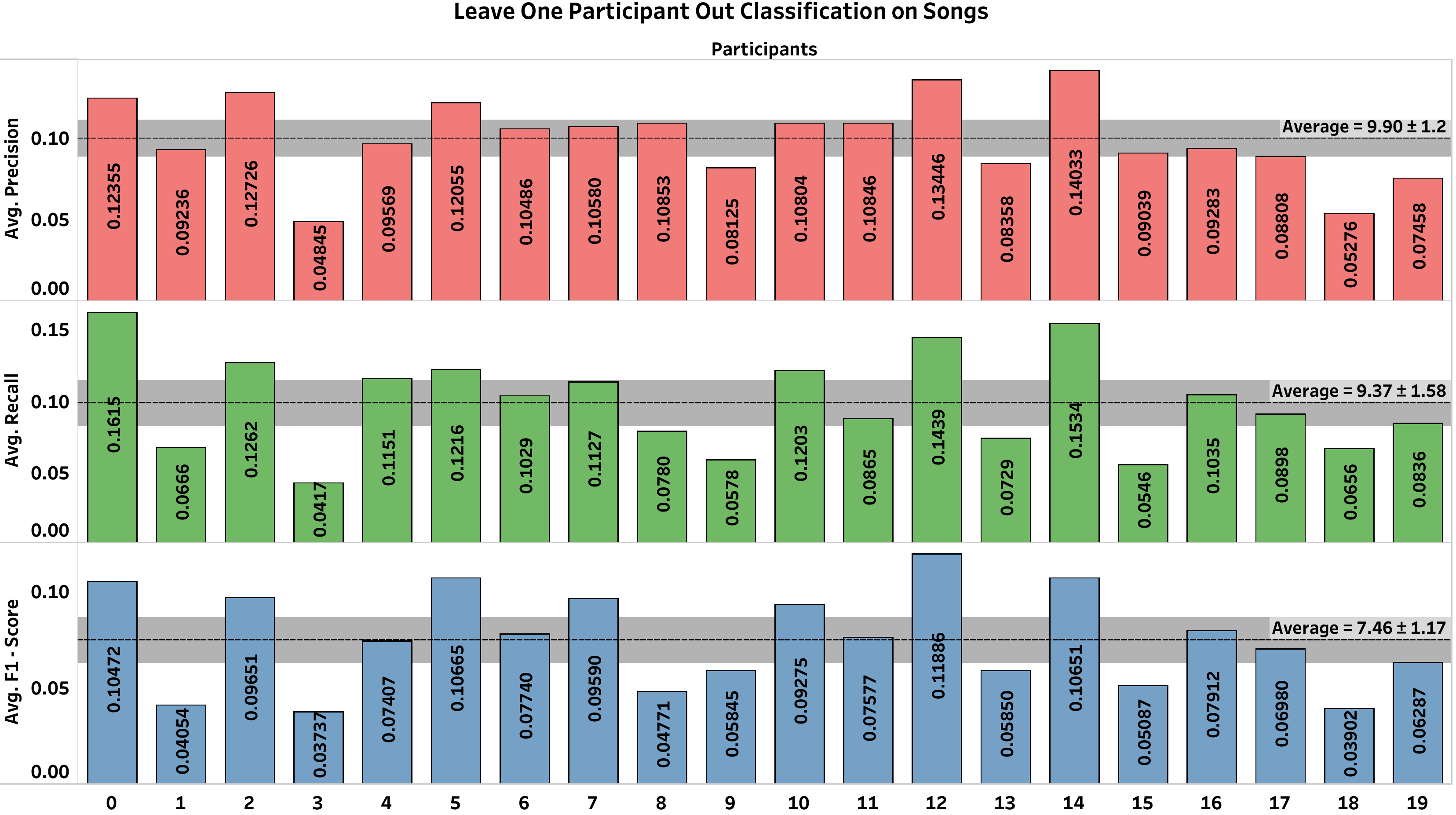}
  \caption{Leave-one-user-out Song identification. The dotted line represents average score across all participants, with the gray band denoting the 95\% confidence interval.}
  \label{all}
  \Description{}
\end{figure}

For song identification, we observed an average test accuracy of $92.83\pm0.04$, test precision of $92.87\pm0.02$, test recall of $92.83\pm0.03$ and test F1-score of $92.82\pm0.02$ for ten repetitions of stratified 5-fold validation. However, as seen from Figure~\ref{all}, the CNN model can identify songs only at less-than-chance level (chance-level = 10\%) for leave-one-user-out cross-validation. This means that the CNN model is unable to learn generalized neural embeddings for discriminating songs. Overall, the model performs poorly when no sample of the target (test) user is part of the training set. These observations have been echoed in~\cite{sonawane2021guessthemusic} for song classification, where poor results were obtained after randomly omitting five participants from a group of twenty. 

\section{Song Identification incorporating Enjoyment Ratings}
\begin{figure}[h]
      {\includegraphics[scale=0.35]{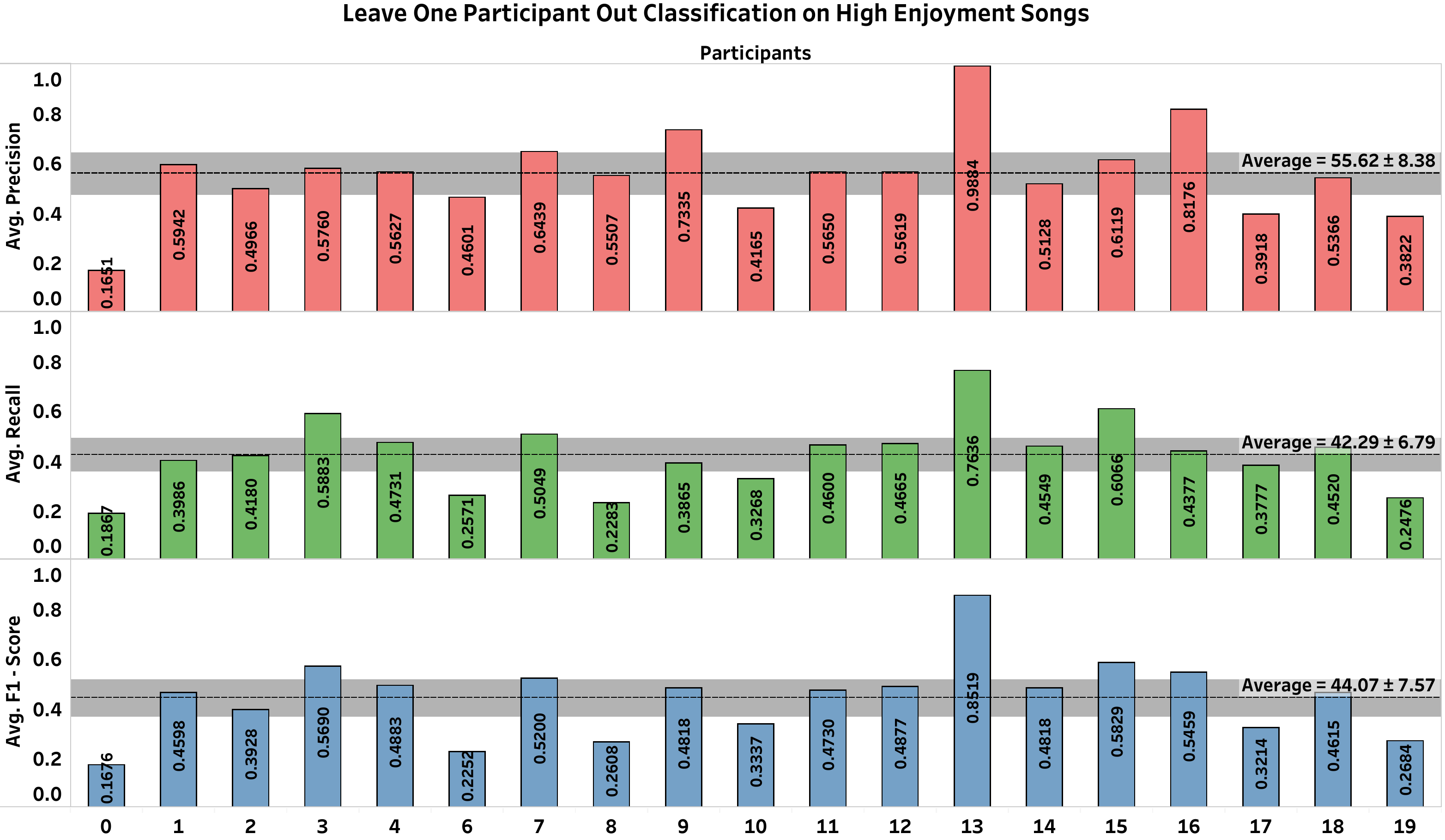}}
\vspace*{\fill}
\hspace*{\fill}
      {\includegraphics[scale=0.50]{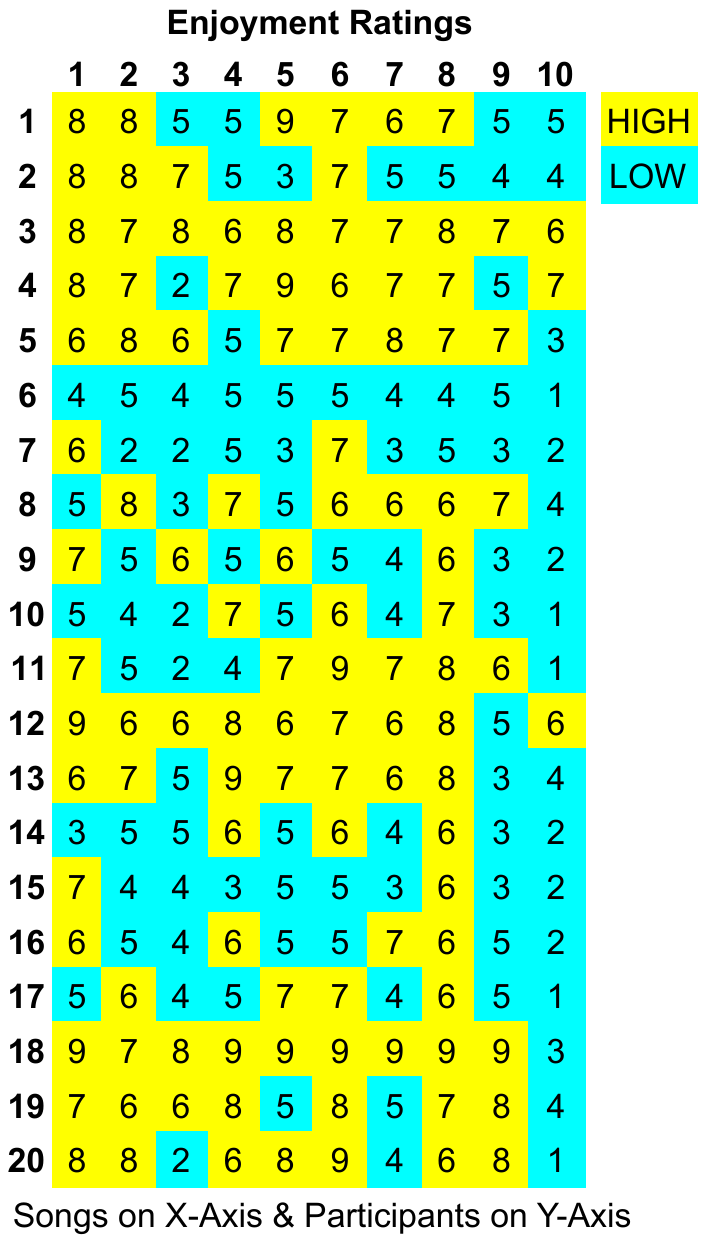}}
\caption{(a) Leave-one-user-out Song identification conditioned on high enjoyment. The dotted line represents average score across all participants, with the gray band denoting the 95\% confidence interval. (b) Enjoyment ratings with binary class labels.}
\label{hist_heat_high}
\end{figure}

Enjoyment denotes a positive affective state that arises when an individual engages in a satisfying activity. The song classification experiments convey user-dependent EEG encodings generated during music appreciation, indicating that different brains listen to the same song differently. Furthermore, brain responses are known to be heavily influenced by the user's affective state, including enjoyment. So we set out to investigate if there is any difference when song classification is conditioned on enjoyment ratings, available as part of the NMED-T dataset. We hypothesized that all participants who expressed low/high enjoyment for a song, must have similar brain encodings. During dataset acquisition, participants rated their enjoyment on a scale of 1-9 (Fig.~\ref{hist_heat_high}(b)). We set a threshold of five to dichotomize enjoyment ratings; for each user, we chose songs with ratings greater than five and assigned them to the \textit{high} enjoyment class, while all other songs were assigned to the \textit{low} enjoyment class. Across all song samples, 91 samples were associated with high enjoyment, while 109 were associated with low enjoyment. 

\begin{figure}[h]
  \centering
  \includegraphics[scale=0.35]{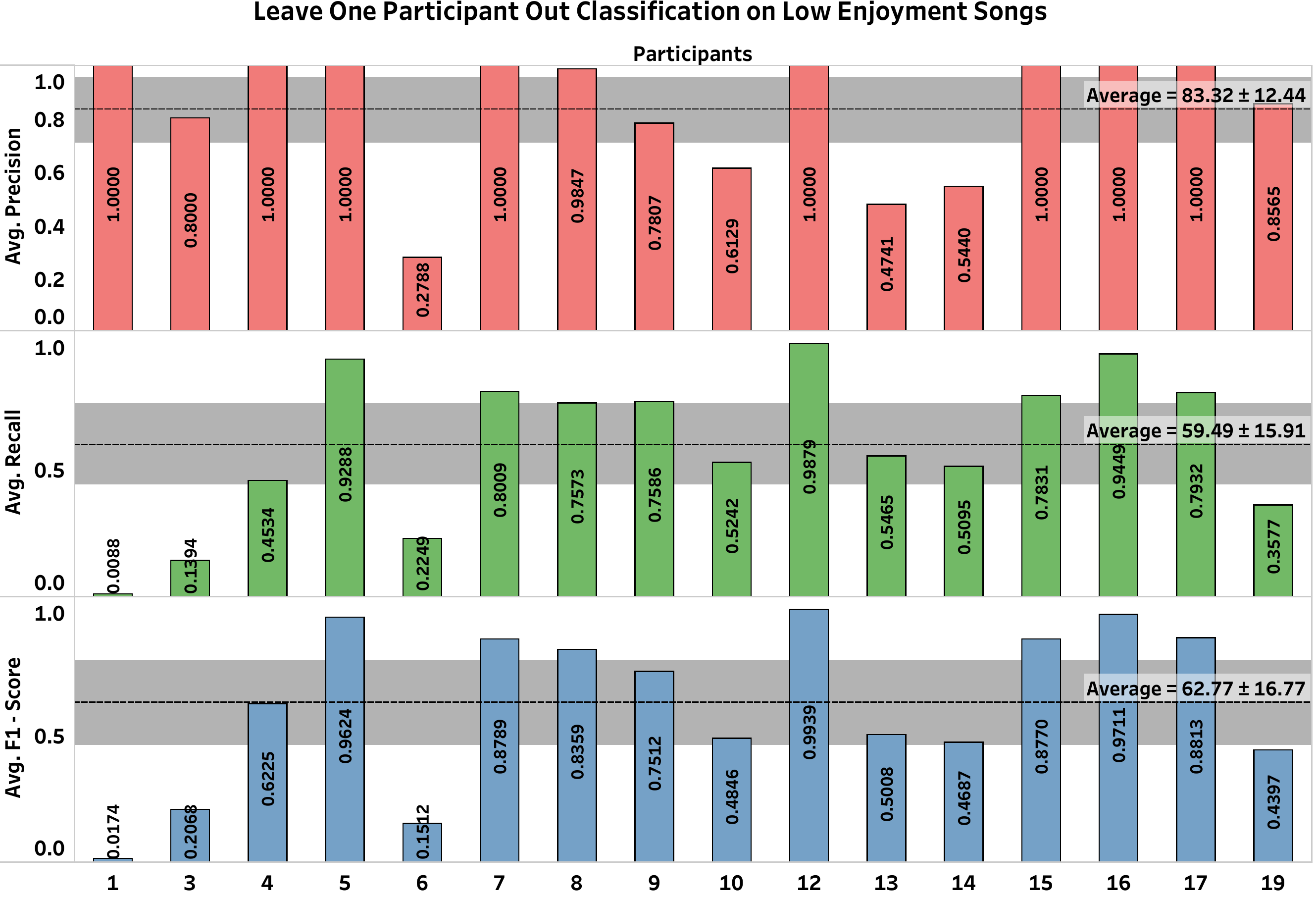}
  \caption{Leave-one-user-out Song identification conditioned on low enjoyment. The dotted line represents average score across all participants, with the gray band denoting the 95\% confidence interval.}
  \label{low}
  
\end{figure}

We began by examining how well the CNN model recognized \textit{high} and \textit{low} enjoyment. Applying ten repetitions of stratified 5-fold cross-validation, the model achieved $97.94\pm0.01$ accuracy, $97.96\pm0.02$ precision, $97.95\pm0.01$ recall, and $97.95\pm0.02$ F1-score. This result revealed that the CNN model could efficiently classify EEG encodings into binary enjoyment classes. Following this, we retrained the CNN for song classification in the leave-one-user-out context, conditioned on high or low enjoyment. In the high enjoyment set, we found an average precision, recall, and F1-score of $55.62\pm8.38$, $42.29\pm6.79$, and $44.07\pm7.57$, respectively as shown in Figure \ref{hist_heat_high} (a), while in the low enjoyment set, the average precision, recall, and F1-score were $83.32\pm12.44$, $59.49\pm15.91$, and $62.77\pm16.77$ as shown in Figure~\ref{low}.

These results convey that while there may be significant neural encoding differences across users, segmenting EEG data based on enjoyment ratings achieves considerably better performance, or makes the CNN model much more generalizable. That EEG similarities across users can be captured better upon segmenting with respect to enjoyment implies that song enjoyment has an key role to play in its neural encoding. Our results suggest that additional attributes need to be accounted for to perform efficient music recognition from neural signals.     

\section{Conclusion}
EEG-based song recognition would facilitate applications such as song retrieval/recommender systems. Based on empirical evidence, we attempt to address the complexity of music classification and report the significance of enjoyment. Despite limited data, we opted for a CNN owing to its ability to efficiently learn from high-dimensional EEG data, and our results convey minimal overfitting. For song recognition, the CNN model performs well when the training data includes the target user's samples. However, model performance dips when the training data includes no samples of the target user. This dip is alleviated when the data is partitioned based on \textit{enjoyment} ratings. 
Our results elucidate the formation of similar neural patterns in users who experience similar enjoyment. Our findings encourage further investigation into identifying additional factors that can be used to classify neural encodings in response to naturalistic music. One potential route would be to investigate the impact of perceptual attributes such as familiarity, engagement, etc., to enable accurate subject-independent song recognition. 

\section*{Acknowledgements}
We thank Science and Engineering Research Board (SERB) and PlayPower Labs for supporting the Prime Minister’s Research Fellowship (PMRF) awarded to Pankaj Pandey and FICCI for facilitating this PMRF.

\bibliographystyle{ACM-Reference-Format}
\bibliography{sample-base}

\end{document}